\documentclass[10pt]{iopart}

\usepackage{graphicx}
\usepackage{xcolor}
\usepackage{amsmath}

\begin{document}

\title[]{\Large Gyrokinetic investigation of toroidal Alfv\'en eigenmode (TAE) turbulence}

\author{Ajay C. J.$^1$, Ben McMillan$^1$, Arkaprava Bokshi$^2$, Alessandro di Siena$^3$, M. J. Pueschel$^{4,5,6}$, Juan Ruiz Ruiz$^7$}

\address{$^1$Centre for Fusion, Space and Astrophysics, Department of Physics, University of Warwick, CV4 7AL, Coventry, UK\\
$^2$York Plasma Institute, University of York, Heslington, York YO10 5DD, UK\\
$^3$Max Planck Institute for Plasma Physics, Boltzmannstr.2, 85748 Garching, Germany\\

$^5$Dutch Institute for Fundamental Energy Research, 5612 AJ Eindhoven, The Netherlands\\
$^5$Eindhoven University of Technology, 5600 MB Eindhoven, The Netherlands \\
$^6$Department of Physics \& Astronomy, Ruhr-Universit{\"a}t Bochum, 44780 Bochum, Germany\\
$^7$Rudolf Peierls Centre for Theoretical Physics, University of Oxford, Oxford OX1 3PU, UK}
\ead{trax.42@hotmail.com}
\vspace{10pt}
\begin{indented}
\item[]December 2023
\end{indented}

\begin{abstract}
Toroidal Alfv\'en eigenmodes (TAEs) can transport fusion-born energetic particles out of the plasma volume, thereby decreasing plasma self-heating efficiency and possibly damaging reactor walls. Therefore, understanding TAE destabilisation and identifying saturation mechanisms is crucial to achieving burning plasma. 
While TAEs have been studies extensively in the past using kinetic-MHD codes, here a fully gyrokinetic study is employed which allows for additional physics. 
 In the case studied, the primary drive mechanism is identified as the resonance between the magnetic drifts and the TAE, and this is seen to be disrupted by equilibrium flow shear which can stabilize the mode by rotating it in the the poloidal plane.
 It is found that zonal flows do not play a significant role in the saturation of these TAEs, and there are no saturation mechanisms present in the local gyrokinetic picture that are able to saturate the mode at physically relevant transport levels in the case of TAE-only turbulence.
 Instead, we confirm that the global profile flattening of fast-ion density is the key saturation mechanism.
 The nonlinear excitation of TAE travelling along the electron diamagnetic direction and its beating with the ion diamagnetic TAE, resulting in large amplitude oscillations that may help detect TAEs more easily in tokamaks, is also reported.
\end{abstract}

%
%
%
%
\ioptwocol

\section{Introduction}

Alfv\'enic modes such as Toroidal Alfv\'en Eigenmodes (TAEs)~\cite{Cheng1986,Fu1989} can be detrimental to tokamak operation as they can transport significant fast-ion population out of the plasma. In addition to decreasing the thermonuclear alpha population necessary for self-burning, the expelled fast-ions may strike and damage the reactor walls. Hence, a proper understanding of TAE destabilisation, damping and saturation  mechanisms is crucial.

Many damping mechanisms such as continuum damping~\cite{Zonca1992,Berk1992}, landau damping~\cite{Fu1989,Betti1992}, radiative damping~\cite{Mett1992}, finite-orbit width effects~\cite{Fulop1996} etc. can stabilise TAEs linearly. However, once destabilised, usually by the radial gradient of fast-ion pressure, the level at which TAEs saturate nonlinearly becomes relevant for transport prediction. Fast-ion profile relaxation is one of the obvious ways by which it saturates. Other possible saturation pathways may include trapping of fast-ions by finite amplitude TAEs~\cite{Todo1995}, nonlinear coupling to other micro/MHD instabilities or zonal modes.


Kinetic treatment of fast-ions is necessary to model the fast-ion destabilisation of TAEs. Therefore, hybrid codes~\cite{Todo1998,Fu2006,Wang2011} where the background species are treated using MHD equations and the fast-ions are treated kinetically are usually used. However these codes lack the ability to model certain physics that influence mode saturation, such as the nonlinear coupling~\cite{Chen2011} between short scale structures associated with the continuum, wave-particle interaction necessary for ion induced scattering of TAEs~\cite{Hahm1995}, trapped particle effects that play an important role in zonal flow structure generation~\cite{Chen2012}, etc., and hence a full gyrokinetic formalism may be necessary~\cite{Qiu2023}. 


In this paper, we explore TAE-only turbulence with the help of the gyrokinetic code \textsc{Gene}~\cite{GENE1,GENE2} to study the destabilisation process and identify the saturation mechanisms. Other works~\cite{DiSiena2018,DiSiena2019,Biancalani2021} have looked at coupled scenarios where complex interactions between drift waves and Alfv\'enic modes have been found to affect transport properties. Hence, the relevance of studying the simplest case of TAE-only turbulence can be understood in context. 

We show the resonance of the magnetic (curvature and $\nabla B$) drift of fast-ions with the TAE in velocity space which facilitates the destabilisation mechanism. The poloidal localisation of the destabilisation region is shown and the stabilisation of TAE by $E\times B$ flow shear in medium-high flow-shear tokamaks is discussed. TAEs are found to excite significant zonal flows but they are not found to play a role in saturation. Global flattening of the profile gradient is identified as the most important saturation mechanism with the help of a global code and the inadequacy of the local code to predict TAE transport quantitatively is stressed.

This paper is organised as follows. In section \ref{SecSimSetup}, the simulation setup is explained, followed by linear results on the mode structure and the destabilisation mechanism in section \ref{SecLinear}. In section \ref{SecNonlinear}, the saturation mechanisms are explored with the help of nonlinear simulations, and finally the conclusions are presented in section \ref{SecConclusions}.

\section{Simulation set-up}\label{SecSimSetup}
The \textsc{GENE} simulations use a field-aligned coordinate system~\cite{Beer1995} where $x$ is the radial coordinate, $y$ the binormal coordinate and $z$ the parallel coordinate. Parallel velocity $v_\parallel$ and magnetic moment $\mu$ are the velocity space coordinates. 

Two versions of the code, \emph{local}~\cite{Scott1998} and \emph{global}~\cite{GENE2,GoerlerPhD} are used. The local version assumes radially constant background gradients, making the set-up and the physics simpler. The global version allows radially varying background gradients, and as will be discussed in detail in section~\ref{SecNonlinearglobal}, is necessary to predict the flux levels accurately in the case of TAEs.

Given that the primary motivation behind this work is to study the fundamentals of TAE destabilisation and saturation, the parameters are chosen such that a simple TAE-only dominated regime is obtained in our simulations rather than one taken directly from experiments which might be a mix of many instabilities. The well studied Cyclone Base Case (CBC)~\cite{Dimits2000} parameters are chosen and then tweaked. A fast-ion species (denoted by subscript $f$) is added along with electrons ($e$) and thermal ions ($i$). Concentric circular flux-surface geometry \cite{Lapillonne2009} is considered with an inverse aspect ratio of $\epsilon=x_0/R=0.18$. Mass ratios of $m_f/m_i=1$ and $m_e/m_i=2.5\times 10^{-4}$, density ratios of $n_f/n_e=0.01$ and $n_i/n_e=0.99$, temperature ratios of $T_f/T_i=100$ and $T_e/T_i=1$, charge ratios of $q_f/q_i=1$ and $q_e/q_i=-1$, and a normalised electron pressure of $\beta_e=0.001$ are considered. Collisions and $\delta B_\parallel$ fluctuations are not included. Numerical hyperdiffusion~\cite{Pueschel2010_2} is included with coefficients $D_z=3.0$ and $D_v=1.0$ along the parallel and parallel velocity coordinates respectively. 

In the \emph{local} simulations, the inverse of the density or temperature background gradient scale lengths, normalized to the major radius $R$, are $R/L_{N,f}=25.0$, $R/L_{N,i}=0$, $R/L_{N,e}=0.25$, $R/L_{T,f}=0$, $R/L_{T,i}=0$ and $R/L_{T,e}=0$. A safety factor of $q_0=1.4$ and magnetic shear of $\hat{s}=0.8$ are considered. The default nonlinear simulation has a minimum binormal wavenumber of $k_{y,\min}\rho_i=0.01$ ($L_y=628.3\rho_i$) where $\rho_i$ is the thermal ion Larmor radius, radial box size of $L_x=125.0\rho_i$ and velocity space box sizes of $L_{v\parallel}=3\sqrt{2}v_{th,i}$ ($v_{th,i}=\sqrt{T_{0,i}/m_i}$) and $L_\mu=12 T_{0,i}/B_{0,{\rm axis}}$. The grid resolutions are $N_x\times N_{y}\times N_z\times N_{v_\parallel}\times N_{\mu} = 64\times 20\times 32\times 36 \times 12$. A larger radial box-size simulation with $L_x=1000\rho_i$ and $N_x=512$ is also used.

The \emph{global} simulation, centered at $x_0=0.5a$, where $a$ is the tokamak minor radius, spans a radial width of $L_x=62.5\rho_i$, with $\rho^\star=\rho_i/a=0.0036$. A quadratic  q-profile of the form $q(x)=0.84+2.24(x/a)^2$ is considered such that both the safety factor and magnetic shear match that of the local simulations at the center. For practical reasons in \textsc{GENE} related to satisfying quasi-neutrality, an extra electron species labelled $en$ is included such that $n_e=0.99n_0$, $n_{en}=0.01n_0$, $n_i=0.99n_0$ and $n_f=0.01n_0$. The radial background temperature and density profiles are of the form $A_s={\rm exp}[- \kappa_{A,s} \ \epsilon\  \Delta A_s\ {\rm tanh}((x-x_0)/(a\Delta A_s))]$ where $A_s$ represents the temperature or density of species $s$, $\kappa_{A,s}$ denotes the peak gradient and $\Delta A_s$ denotes the radial width of the gradient profile; $\kappa_{n,e}=0$, $\kappa_{n,en}=95$, $\kappa_{n,i}=0$, $\kappa_{n,f}=95$, $\kappa_{T,e}=\kappa_{T,en}=\kappa_{T,i}=\kappa_{T,f}=0$, and $\Delta n_{en}=\Delta n_{f}=0.05$. The minimum binormal wavenumber is $k_{y,\min}\rho_i=0.02$ corresponding to minimum toroidal mode number $n_0=2$ and  the numerical resolutions are $N_x\times N_{y}\times N_z\times N_{v_\parallel}\times N_{\mu} = 48\times 10\times 32\times 36 \times 12$. Krook heat and particle sources (see Ref.~\cite{Lapillonne2011_2} for details) are also employed with a source rate of $\gamma_h=\gamma_p=0.015 v_{th,i}/R$ which is approximately an order of magnitude lower than the maximum growth rate so that the time scale on which the source rate changes the profile is smaller than the characteristic time of turbulence.

\section{Investigating the destabilisation mechanism using linear simulations}\label{SecLinear}
Linear simulations are used to probe the details of the linear TAE mode including its mode structure, the destabilisation mechanism and ballooning angle dependence. The local results are given in section~\ref{SecLocLinear} followed by global results in section~\ref{SecGlobalLinear}.

\subsection{Local linear results}\label{SecLocLinear}

\subsubsection{Growth rate, frequency and mode structure.}
\begin{figure}[h]     		
\includegraphics[scale=0.7]{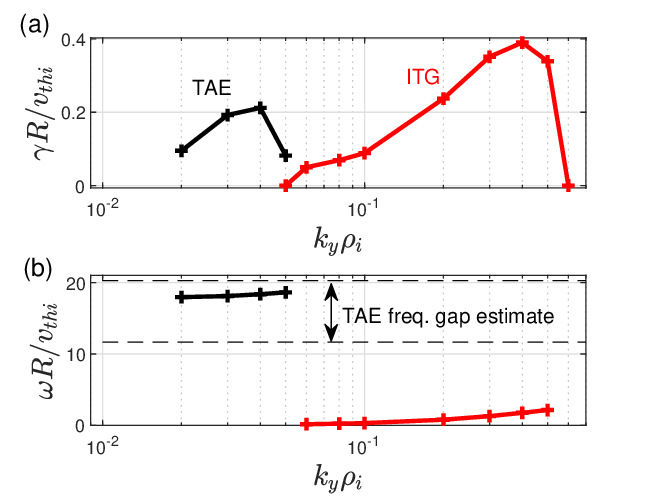}
\caption{(a) Growth rate and (b) frequency as a function of binormal wavenumber $k_y$ in linear local simulations. TAE and ITG modes are shown in black and red colors respectively. }
\label{FigLinky}
\end{figure}

TAEs are found to be unstable at wavenumbers lower than those where Ion Temperature Gradient (ITG) modes are most unstable in a standard CBC scenario. See figure~\ref{FigLinky} where the growth rates and frequencies are plotted as a function of the binormal wavenumber. The TAE mode frequencies clearly lie within the analytic TAE frequency gap that is estimated as $3\epsilon|k_\parallel v_A|$ following Fu and Van Dam \cite{Fu1989}. The gap is centered at the mid-frequency $\omega_{\rm TAE}=k_\parallel v_a$. Here, $k_\parallel=1/(2Rq_0)$ is the parallel wavenumber at (the half-mode-rational surface of) intersection of modes with poloidal mode numbers $m$ and $m+1$, and $v_A=B_0/\sqrt{\mu_0\sum_sn_sm_s}$ is the Alfv\'en speed \cite{Heidbrink2008}.

\begin{figure}[h]     		
\includegraphics[scale=0.78]{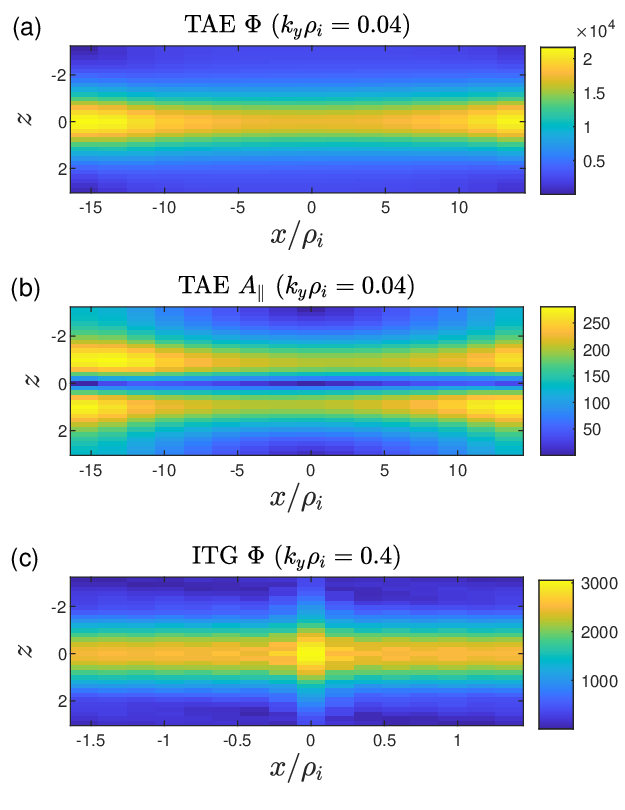}
\caption{(a) Electrostatic potential $\Phi$ and (b) parallel component $A_\parallel$ of the vector potential plotted on the $x-z$ plane for the TAE mode with $k_y\rho_i=0.04$. (c) $\Phi$ of the ITG mode with $k_y\rho_i=0.4$.}
\label{FigLinCont}
\end{figure} 

The TAE mode has peak amplitude near the half-rational surface, as can be observed in figure~\ref{FigLinCont}(a and b) where the electrostatic potential and vector potential respectively are plotted on the $x-z$ plane for the most unstable TAE having $k_y\rho_i=0.04$. Note that in linear flux-tube \textsc{GENE} simulations, the mode rational surfaces are located at the middle of the radial domain ($x=0$) and the half-mode-rational surfaces are located at the edges~\cite{AjayCJPhD}. In contrast, the ITG mode simulated with kinetic electrons has peak amplitude near the mode rational surface~\cite{Dominski2015,AjayCJ2020} as shown in figure~\ref{FigLinCont}(c).

\begin{figure}[h]     		
\includegraphics[scale=0.7]{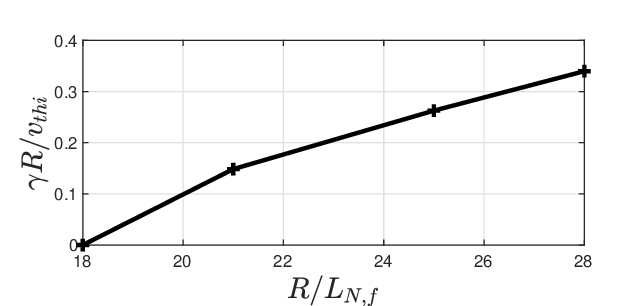}
    \caption{Growth rate of TAE ($k_y\rho_i=0.04$) mode plotted as s function of normalised fast-ion density gradient $\omega_{n,f}=R/L_{N,f}$.}
\label{FigLinLn}
\end{figure}

The default normalised fast-ion density gradient $R/L_{N,f}=25.0$ is well above the marginal value $R/L_{N,f}=18.0$ as can be seen in figure~\ref{FigLinLn}.

\subsubsection{Investigating the destabilisation mechanism.}
To investigate the basic destabilisation mechanism of TAEs, the method employed by Di Siena et al.~\cite{DiSiena2019} to study fast-ion stabilisation of ITG is used. This involves individually removing terms from the gyrokinetic equation responsible for each of the main physical process. 

The relevant gyrokinetic equation in the GENE field aligned coordinates, assuming Maxwellian background distribution function, reads
\begin{equation}
\begin{split}
&-\frac{\partial}{\partial t} f_{1,j}
=
\\
&\underbrace{\frac{1}{\mathcal{C}}
\left[ \frac{d {\rm ln} n_{o,j}}{dx}
+ \left(
\frac{m_jv_\parallel^2}{2T_{0,j}} + \frac{\mu B_0}{T_{0,j}} - \frac{3}{2}
\right) \frac{d {\rm ln} T_{o,j}}{dx}
\right] \frac{\partial \bar{\Phi}}{\partial y} f_{0,j}}_{\rm background\ drive}
\\
&+
\underbrace{
\frac{\mu B_0+m_jv_{\parallel}^2}{m_j\Omega_j}
\mathcal{K}_x\Gamma_{x,j}
}_{\rm radial\ magn.\ drift}
+ 
\underbrace{
\frac{\mu B_0+m_jv_{\parallel}^2}{m_j\Omega_j}
\mathcal{K}_y\Gamma_{y,j}
}_{\rm binormal\ mag.\ drift} 
\\
&+
\underbrace{\frac{\mathcal{C}v_{\parallel}}{B_0J^{xyz}}\Gamma_{z,j}}_{\rm parallel\ advection} -
\underbrace{\frac{\mathcal{C}\mu}{m_jB_0J^{xyz}}\frac{\partial B_0}{\partial z}\frac{\partial f_{1j}}{\partial v_{\parallel}}}_{\rm trapping}
\end{split}
\label{GKEq}
\end{equation}
where, $\mathcal{C}=B_0/|\nabla x\times\nabla y|$, $\Gamma_{\alpha, j}=
\frac{\partial f_{1j}}{\partial \alpha} -
\frac{q_j}{m_jv_\parallel}\frac{\partial \bar{\Phi}_1}{\partial \alpha}\frac{\partial f_{0j}}{\partial v_\parallel},
\mathcal{K}_x=
-\frac{1}{\mathcal{C}}\frac{\gamma_2}{\gamma_1}\frac{\partial B_0}{\partial z},
\mathcal{K}_y=
\frac{1}{\mathcal{C}}
\left(
\frac{\partial B_0}{\partial x} -
\frac{\gamma_3}{\gamma_1}\frac{\partial B_0}{\partial z}
\right)$, $\gamma_1=g^{xx}g^{yy}-(g^{xy})^2$, $\gamma_2=g^{xx}g^{yz}-g^{xy}g^{xz}$, $\gamma_3=g^{xy}g^{yz}-g^{yy}g^{xz}$, $g^{\alpha\beta}=\nabla\alpha\cdot\nabla\beta$, with $\alpha,\beta= x,y,z,$ and $\bar{\Phi}$ denotes the gyro-averaged $\Phi$. In the above equation, the neoclassical and nonlinear terms have been excluded since they don't affect the evolution of $f_{1,j}$ in the linear flux-tube limit.

The background drive term determines the evolution of perturbed distribution function due to spatial gradients of the background distribution, and appears as a source term. Other terms appear as phase space advection velocity, such as the magnetic (curvature and $\nabla B$) drifts and the parallel motion and trapping terms. The magnetic drifts are further separated into their radial and binormal components. The results are summarized in table~\ref{TabGKrem}. Numerical hyperdiffusion is not included in these simulations for simplicity and hence the growth rates are slightly off from those in figures~\ref{FigLinky} and \ref{FigLinLn}.

\begin{table}
\begin{center}
\begin{tabular}{ c c}
 Terms removed & $\gamma R/v_{th,i}$ \\ 
 \hline
 None (Original) & 0.33\\  
 Background drive & 0 \\
 Par. adv. and trap. & 0.48 \\
 x-curvature & 0.31 \\
 y-curvature & 0
\end{tabular}
\end{center}
\caption{Growth rate of the $k_y\rho_i=0.04$ TAE modes when terms in the gyrokinetic equation are removed.}
\label{TabGKrem}
\end{table}

As expected, removing the fast-ion background drive term stabilises the TAE. Removing the fast-ion parallel advection and trapping terms leads to higher growth rate, indicating that they provide a net stabilising effect. Radial magnetic drift is found to have little effect on the mode. The most insightful result is the stabilisation of the mode in the absence of fast-ion binormal magnetic drift term, indicating the relevance of this term in TAE destabilisation. 

As the modes are close to ideal MHD, the parallel electric field is close to zero, and the drive from fast ions is mostly due to the (perpendicular) magnetic drifts allowing a $\bf J\cdot E$ energy exchange to the mode, but this also requires background pressure gradients. Therefore, net growth requires that both of these terms be positive. This mechanism also mostly involves drifts in the binormal direction, which is the main way the electric field is oriented.

\begin{figure}[h]     		
\includegraphics[scale=0.7]{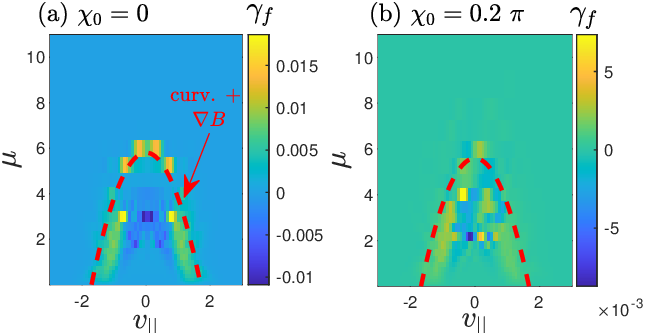}
\caption{Fast-ion free-energy based growth rate $\gamma_f$ plotted on the $v_\parallel-\mu$ velocity plane for ballooning angles (a) $\chi_0=0$ and (b) $\chi_0=0.2\pi$ for the $k_y\rho_i=0.04$ TAE. Red Dashed line indicates the magnetic (curvature and $\nabla B$) drift resonance curve.}
\label{FigLinVel}
\end{figure}

One can obtain better insight on the fast-ion-TAE resonance by visualising the wave-particle energy transfer in the velocity space. For this purpose, a free-energy based diagnosis~\cite{Navarro2011, DiSiena2019} is used. The total system free energy $E$ in the electrostatic case is defined as 
\[
\begin{split}
E = \sum_j{\rm Real}\left[
\int dzd\mu dv_\parallel\frac{\pi B_0 n_0 T_0}{2f_{0,j}} (\Phi^* + f_{1,j}^*) f_{1,j}
\right].
\end{split}
\]
A free-energy based growth rate $\gamma_j$ for each species can be obtained such that
\[
\begin{split}
\gamma = \sum_j\gamma_j
= \frac{1}{E} \sum_j \frac{\partial E_j}{\partial t},
\end{split}
\]
where $\partial E_j/\partial t$ is essentially obtained by multiplying the gyrokinetic equation (\ref{GKEq}) with $(\Phi^* + f_{1,j}^*)$. The fast-ion growth rate is plotted on the $v_\parallel-\mu$ velocity space in figure~\ref{FigLinVel}(a). 

Considering only the background drive and magnetic drift terms in the gyrokinetic equation for fast-ions, and assuming perturbations (\emph{i.e.} $\Phi$ and $f_{1,f}$) to be of the form exp$[-i\omega t+ik_xx+ik_yy]$ in Fourier space, the perturbed distribution function $f_{1,f}$ can be expressed as a fraction where the denominator contains the resonance between the real frequency $\omega_r$ of the wave and the fast-ion magnetic drift frequency, which can be expressed as

\begin{equation}
\omega_r
= -\frac{\mu B+m_fv_\parallel^2}{q_fB}
\kappa_yk_y.
\end{equation}
For more details, see equations (2-8) in reference~\cite{DiSiena2019}. In figure~\ref{FigLinVel}, this resonance curve is denoted by the red dashed line. One can observe significant fast ion growth rate contribution overlapping with the resonance curve in figure~\ref{FigLinVel}(a), confirming that indeed magnetic drifts play a crucial role in the destabilisation mechanism.

The significant contributions at $\mu\simeq 3$ inside the resonance curve in figure~\ref{FigLinVel}(a) result from the parallel advection and trapping contribution, which is confirmed by their absence in a similar figure (not shown) corresponding to the simulation where the parallel advection and trapping terms are removed. These terms have a net stabilising effect as confirmed in table~\ref{TabGKrem}. 

\subsubsection{Ballooning angle dependence.}\label{SubSecBal}
The curvature and $\nabla B$ magnetic drifts, which are primarily oriented along the vertical direction ($\approx \bf B_0\times\nabla B_0$) can resonantly drive TAEs only at those poloidal locations where the TAE phase fronts (oriented approximately along the poloidal direction) have significant component along the same vertical direction. More precisely, these correspond to those poloidal locations where the geometric factor $\kappa_y$ is negative. 

\begin{figure}[h]     		
\includegraphics[scale=0.7]{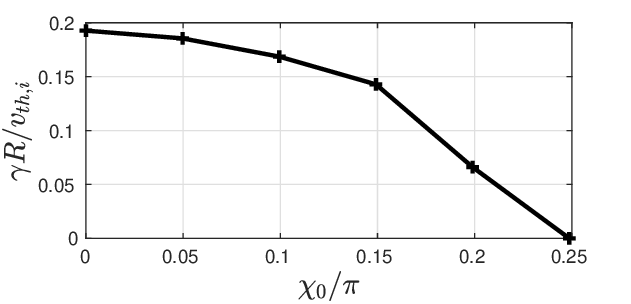}
\caption{Linear growth rate of the $k_y\rho_i=0.04$ TAE mode as a function of ballooning angle $\chi_0$.}
\label{FigLinBal}
\end{figure}

In the default case of zero ballooning angle when the mode is localised at the outboard mid-plane and both the phase fronts and the magnetic drifts are oriented along the same vertical direction, maximum destabilisation happens.
This is verified in figure~\ref{FigLinBal}. The growth rate is seen to decrease with increasing ballooning angle, fully stabilising at $\chi_o=0.25 \pi$. Further evidence can be seen in figure~\ref{FigLinVel}(b) for the case with a finite ballooning angle of $\chi_0=0.2\pi$, where the free-energy based growth rate contributions along the resonance curve can be seen to be mostly destroyed.

\subsection{Global linear results}\label{SecGlobalLinear}
\begin{figure}[h]     		
\includegraphics[scale=0.7]{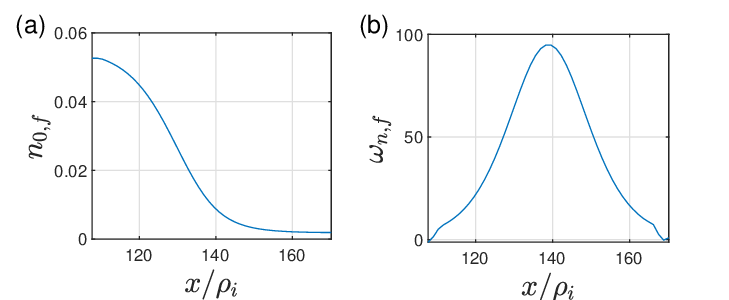}
\caption{Background fast-ion (a) density $n_{0,f}$ and (b) normalised logarithmic gradient $\omega_{n,f}$ plotted as function of the radius in global simulations.}
\label{FigLinBG}
\end{figure}

\begin{figure}[h]     		
\includegraphics[scale=0.7]{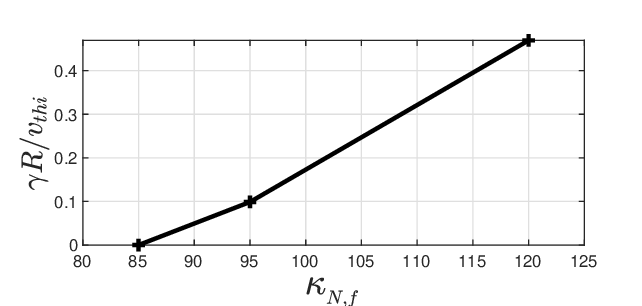}
\caption{Growth rate of the global TAE mode as a function of normalised logarithmic fast-ion gradient $\omega_{n,f}$.}
\label{FigLinGlogvsomnf}
\end{figure}

\begin{figure}[h]     		
\includegraphics[scale=0.78]{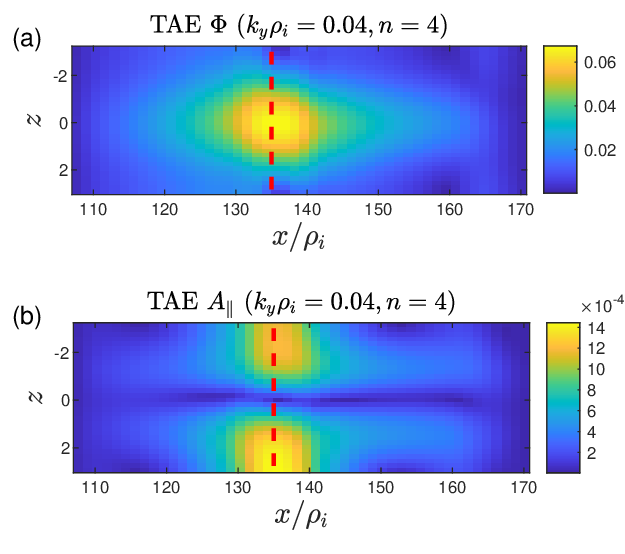}
\caption{(a) $\Phi$ and (b) $A_\parallel$ plotted on the $x-z$ plane for the global TAE mode with $k_y\rho_i=0.04$, $n=4$. Red dashed line indicates the $q=1.375$ surface, which is the half-rational surface of intersection of the $m=5$ and $m=6$ poloidal modes for $n=4$.}
\label{FigLinGlobalCont}
\end{figure}

The fast ion density profile used in global simulations is shown in figure~\ref{FigLinBG}. In contrast to local simulations, a much higher critical gradient is necessary to destabilise TAEs in global simulations. The growth rate as a function of the peak fast-ion density gradient $\kappa_{nf}$ is plotted in figure~\ref{FigLinGlogvsomnf} for the only unstable mode with $k_y\rho_i=0.04$ ($n=4$); modes with other $k_y$'s are stable. The linear mode structure, shown in figure~\ref{FigLinGlobalCont}, peaks near $x=135\rho_i=0.48a$ corresponding to safety factor $q=1.375$, which is the half-rational surface of intersection of the $m=5$ and $m=6$ poloidal modes for $n=4$. Various global effects can explain the higher pressure gradients required for destabilisation. For instance, the fast particle drifts may be comparable to the width of the strong pressure gradient region and so the effective drive may be smoothed out. Profile shearing, reduced TAE continuum gap closer to the magnetic axis (gap width is proportional to aspect ratio) etc. are other possibilities.

\section{Investigating saturation mechanisms using nonlinear simulations}\label{SecNonlinear}

\subsection{Nonlinear local results}\label{SecLocNonlinear}

\begin{figure}[b]     		
\includegraphics[scale=0.7]{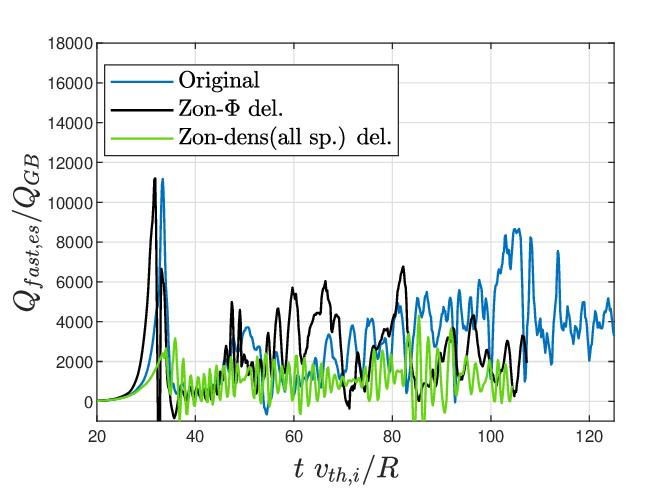}
\caption{Fast-ion electrostatic heat flux normalised by the the gyro-Bohm value plotted as a function of time in local nonlinear simulations with $L_x=125\rho_i$. The original simulation is denoted by blue, and the ones with zonal flow ($\Phi$) and zonal density deleted are denoted by black and green respectively.}
\label{FigNLQvst}
\end{figure}

Local nonlinear simulations of TAE-only turbulence give much higher heat flux levels than the typical heating power available in tokamaks. See figure~\ref{FigNLQvst}, \ref{FigNLglloc}(a) and \ref{Figfreq}(a) where the fast-ion heat flux time-trace for the default case is plotted in blue; the latter figure is plotted with a higher time resolution. Even when the background fast-ion density gradient is decreased closer to the marginal value, the flux-levels still remains unphysically high, for instance, as shown in \ref{FigNLglloc}(b) in magenta for the case with $R/L_{N,f}=20$. The fluxes drop to zero once the linear critical gradient at $R/L_{N,f}=18$ is crossed.

\begin{figure}[h]     		
\includegraphics[scale=0.65]{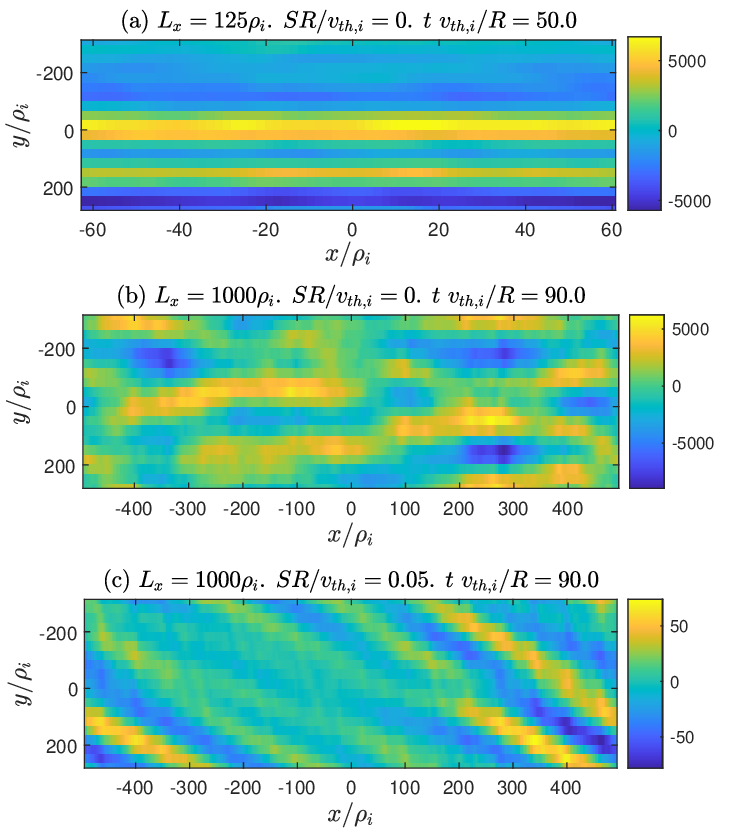}
\caption{Electrostatic potential $\Phi$ plotted on the $x-y$ plane for the case with radial box-size of (a) $L_x=125\rho_i$ and (b) $L_x=1000\rho_i$, both for the zero background shear-flow ($S$) case . (c) $L_x=1000\rho_i$ and $SRv_{th,i}=0.05$.}
\label{FigNLcont}
\end{figure} 

\begin{figure}[b]     		
\includegraphics[scale=0.7]{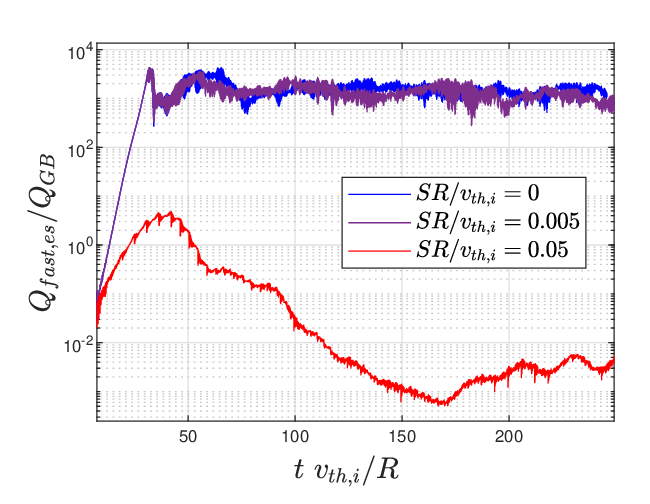}
\caption{Fast-ion electrostatic heat flux normalised by the the gyro-Bohm value plotted as a function of time in local nonlinear simulations with $L_x=1000\rho_i$. The original zero flow-shear simulation is denoted by blue. Purple and red correspond to flow-shear rates of $SRv_{th,i}=0.005$ and $0.05$ respectively.}
\label{FigNLQvstLx1000}
\end{figure}

The local simulations have extremely large streamers extending several hundreds of thermal ion Larmor radii. For the default case with a radial box size of $L_x=125\rho_i$, the streamers extend all through the radial domain as seen in figure~\ref{FigNLcont}(a). The flux-tube formalism assumes that the radial box-size is larger than the characteristic radial length of turbulent eddies. Hence, to contain the eddies within the radial domain, simulations with a much larger radial box-size of $L_x=1000\rho_i (>L_y)$ is required; see figure~\ref{FigNLcont}(b). However, the local (flux-tube) ordering is not well satisfied at such large length scales and a global approach will be more appropriate. These larger radial box size simulations too give unphysically high heat fluxes, as shown in figure~\ref{FigNLQvstLx1000}. Despite the shortcomings of the local model, we can still use these simulations to make certain deductions, as will be discussed in the following.

Given the low wavenumbers of the unstable TAEs, it is unsurprising that the eddies are radially large, however other instabilities, such as microtearing modes for instance, that are unstable at similar wavenumbers have been known to isotropise and saturate to much lower flux levels~\cite{AjayCJ2023}. This indicates that many of the standard saturation mechanisms such as that via zonal flows, local profile flattening at rational surfaces, etc may not be as relevant for TAE turbulence. 

\subsubsection{Role of zonal flows in saturation.} 
Significant zonal flows are driven by TAEs in our simulations, however they do not play an important role in saturation. To confirm this, simulations with artificially deleted zonal flows are carried out. Much higher heat flux would be expected had zonal flows played an important role in saturation~\cite{Lin1998}. However the heat flux-time trace of these simulations, also shown in figure~\ref{FigNLQvst}, are comparable to the original.

\subsubsection{Role of local profile flattening in saturation.} 
Another mechanism by which turbulence can saturate in flux-tube simulations is by locally flattening the profiles at those radial locations of higher diffusivity. This has been demonstrated for the case of microtearing turbulence~\cite{AjayCJ2023} where the mode and the corresponding diffusivity is localised near the mode rational surfaces of each toroidal mode, and hence the zonal $T_e$ perturbations are modified such that the effective (drive) $T_e$ gradient locally flattens significantly at these locations to facilitate saturation. In the case of TAEs, the diffusivity is localised near second order mode rational surfaces, however the corresponding flattening of fast-ion density at these radial positions is insignificant at only at most $3\%$ of the background density gradient as shown in figure~\ref{FigNLglloc}(c). The maximum flattening remains less than the critical gradient even in cases closer to the marginal value. Furthermore, the simulation with artificially deleted zonal density perturbations, so that the local flattening saturation mechanism would be absent, shown in figure~\ref{FigNLQvst} using the green trace, does not show much higher heat flux. These results indicate that local profile flattening mechanism is not effective to saturate TAE turbulence.

\subsubsection{Role of equilibrium flow shear in saturation.}
$E\times B$ equilibrium flow shear can stabilise TAE turbulence by rotating the mode phase fronts in the poloidal plane, from the outboard mid-plane where the modes are maximally unstable to the unfavourable curvature side (see discussion on ballooning angle dependence in subsection~\ref{SubSecBal}). The default case with a radial box size of $L_x=125\rho_i$ has insufficiently resolved $k_x$ (ballooning angle) modes to properly capture the effects of flow shear~\cite{McMillan2019}, and hence $L_x=1000\rho_i$ simulations were used. Simulations with several values of flow-shear rate $S$ were carried out, where $S=-(r_0/q_0)(d\Omega_{\rm tor}/dx)$, with $\Omega_{\rm tor}$ being the toroidal angular velocity. The flow is set to be purely toroidal~\cite{ToldPhD}. The fast-ion heat flux time trace of these simulations is plotted in figure~\ref{FigNLQvstLx1000}. While $SRv_{th,i}=0.005$ has negligible effect, $SR/v_{th,i}= 0.05$  (a realistic value of flow-shear in a typical tokamak) appears to linearly stabilise these TAE modes. However, nonlinear effects could still allow subcritical turbulence~\cite{Casson2009}. In the time period $t \in [0, 60] v_{th,i}/R$ typical Floquet mode behaviour~\cite{WaltzTrans,Dagnelie2019} can be seen for $SR/v_{th,i}= 0.05$. During the initial transient growth phase, the modes are radially aligned and can extract energy from fast ions, followed by damping once they are sufficiently tilted. Figure~\ref{FigNLcont}(c) shows the tilted mode structure at the end of this time period. These results suggest that flow shear could be an important suppression mechanism for TAE-driven transport of fast ions in medium-high flow-shear tokamaks.

\subsection{Profile flattening in nonlinear global simulation}\label{SecNonlinearglobal}
The relaxation of profile gradients is one of the primary mechanisms by which microturbulence transport in general is quenched. However in local simulations, which assume the gradients and the resulting flux to be the same across the radius, such a relaxation is not always possible, especially when the modes are elongated along the radius as already discussed in section~\ref{SecLocNonlinear}. However there are exceptions such as for example in microtearing instability where the diffusivity is extremely confined near low order rationals and a local flattening at these radial positions can be properly captured even in local simulations~\cite{AjayCJ2023}.

\begin{figure}[h]
\hspace{-2ex}
\includegraphics[scale=0.72]{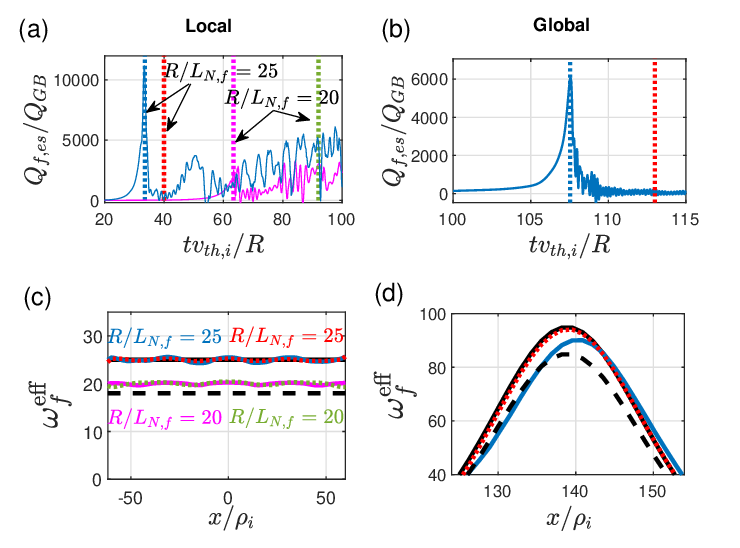}
\caption{Fast-ion heat flux plotted as a function of time in (a) local and (b) global simulations. Effective fast-ion density gradient plotted as a function of the radius in (c) local and (d) global simulations at the time instance indicated by the vertical dotted lines in figures (a) and (b) respectively. Solid black lines denote the original background gradients and dashed lines denote the corresponding critical gradients.}
\label{FigNLglloc}
\end{figure}

In global simulations on the other hand, it is possible to flatten the profiles at those positions where the instability is most unstable, thereby quenching transport. This is observed in TAE global simulations, where unlike in local simulations, the transport is fully quenched, as shown in figure~\ref{FigNLglloc}(b). This emphasises the need to model TAE turbulence via global codes for proper quantitative predictions.

To demonstrate the global profile flattening, the effective fast-ion density gradient $\omega^{\rm eff}_{f}$ is plotted in figures~\ref{FigNLglloc} (c) and (d) for the local and global cases respectively, as a function of the radius, for two specific times. $\omega^{\rm eff}_{f}$ is the total gradient, defined as the sum of the contributions from the background density gradient and the zonal perturbed density gradient, \emph{i.e.}
\begin{equation}
\omega^{\rm eff}_{f}=-\frac{dn_{0,f}/dx}{n_{0,f}/R}-\frac{\langle\partial \delta n_{f}/\partial x\rangle_{yz}}{ n_{0,f}/R}.
\nonumber
\end{equation}
The blue curves in figures~\ref{FigNLglloc} (c) and (d) correspond to that time when the perturbed amplitudes and heat flux peak (denoted by blue vertical dotted lines in figures~\ref{FigNLglloc} (a) and (b)), and there is maximum flattening in the zonal perturbed density gradient. A second curve, red in color, corresponds to a random later time. For the local simulation in figures~\ref{FigNLglloc} (a) and (c), an additional set of plots, denoted in magenta [and green], is added for a case with $R/L_{N,f}=20$, closer to the marginal value. 

For the global simulation, the maximum effective gradient, denoted in blue, can be seen to be below the critical gradient, denoted by the dashed black curve, for a significant portion of the radial domain, sufficient enough to stabilise the mode and drop the corresponding flux-levels to zero. The effective gradient, in fact reduces by a maximum of $15\%$ from the default value denoted in solid black curve. For the corresponding local simulations, there is only a maximum of $3\%$ flattening; the effective gradient remains well above the critical gradient and hence the profile flattening saturation mechanism is not being properly captured.

\subsection{Beating of TAEs travelling along the ion and electron diamagnetic directions}
\begin{figure}[h]
\includegraphics[scale=0.6]{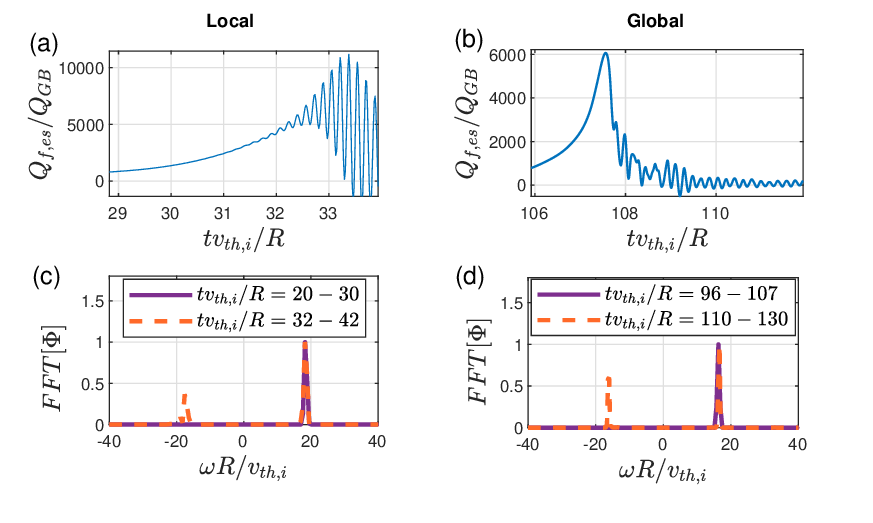}
\caption{Fast-ion heat flux in (a) local and (b) global simulations, zoomed near the linear-nonlinear transition. The fast Fourier transformed electrostatic potential in the linear and nonlinear phases are plotted in violet and orange colors respectively for (c) local and (d) global simulations.}
\label{Figfreq}
\end{figure}

Another interesting observation in TAE turbulence is the intense beating of counter-propagating TAEs in the ion and electron diamagnetic directions. This is illustrated in figure~\ref{Figfreq}. In figures (a) and (b), the zoomed fast-ion heat flux as it transitions from an initial linear phase to the nonlinear phase is plotted for the local and global simulations respectively. In the linear phase, only that TAE travelling along the ion diamagnetic direction is made unstable by the fast-ion magnetic drift resonance. This is confirmed in figures~\ref{Figfreq} (c) and (d) where the fast Fourier transform of the electrostatic potential only has peaks on the positive frequency side corresponding to ion-diamagnetic direction as per \textsc{GENE} convention. Whereas, once the simulation reaches the the nonlinear phase, significant oscillations can be seen on the fast-ion fluxes and other perturbed quantities, having a frequency equal to the most unstable TAE. This essentially results from the nonlinear excitation of electron diamagnetic direction TAE and the subsequent beating with the ion-diamagnetic one. This too is verified in figures~\ref{Figfreq} (c) and (d) where the Fourier transform has peaks both in the positive and negative frequencies. The beating results in high amplitude oscillations in flux surface quantities, which might potentially be directly observed in diagnostics. External magnetic diagnostics are often able to resolve the direction of mode propagation, and this phenomena would be seen as comparable magnitudes in the positive- and negative- propagating mode.

\section{Conclusions}\label{SecConclusions}

Gyrokinetic simulations of TAE turbulence were carried out to study the fast-ion destabilisation and saturation mechanisms. With the help of free-energy based diagnostics, the destabilisation mechanism has been demonstrated to involve the resonance of magnetic drift of fast-ions with the TAE mode. This resonance is more destabilising at the outboard midplane, and hence when the mode gets rotated in the poloidal plane by equilibrium flow shear, it is stabilised. Our simulations predict that TAE turbulence is low or insignificant in medium-high flow-shear tokamaks. Saturation mechanisms such as that via zonal flows and local profile flattening are shown to be less effective and local codes are found to give unphysically transport levels. Global profile flattening on the other hand is shown to be an important saturation mechanism, and therefore the necessity for global codes to quantitatively predict fluxes resulting from TAE turbulence is identified. The nonlinear excitation of electron diamagnetic direction TAEs and their beating with the ion-diamagnetic TAEs, producing large oscillations in observable quantities, is also reported.


\section*{Acknowledgements}
This research was funded in whole or in part by the Engineering and Physical Sciences Research Council (EPSRC), EP/R034737/1. For  the purpose of open access, the author has  applied a Creative Commons Attribution (CC BY) licence to any Author Accepted Manuscript version arising from this submission. We acknowledge the CINECA award under the ISCRA initiative, for the availability of high performance computing resources and support. A. Bokshi would like to acknowledge funding through UKRI Grants EP/R034737/1 and EP/X035336/1, allowing time and access to the ARCHER2 UK National Supercomputing Service  (https://www.archer2.ac.uk).

\section*{References}
\bibliography{TAE_paper} 

\begin{thebibliography}{10}

\bibitem{Cheng1986}
C.~Z. Cheng and M.~S. Chance.
\newblock {Low‐n shear Alfvén spectra in axisymmetric toroidal plasmas}.
\newblock {\em The Physics of Fluids}, 29(11):3695--3701, 11 1986.

\bibitem{Fu1989}
G.~Y. Fu and J.~W. Van~Dam.
\newblock {Excitation of the toroidicity‐induced shear Alfvén eigenmode by
  fusion alpha particles in an ignited tokamak}.
\newblock {\em Physics of Fluids B: Plasma Physics}, 1(10):1949--1952, 10 1989.

\bibitem{Zonca1992}
Fulvio Zonca and Liu Chen.
\newblock Resonant damping of toroidicity-induced shear-alfv\'en eigenmodes in
  tokamaks.
\newblock {\em Phys. Rev. Lett.}, 68:592--595, Feb 1992.

\bibitem{Berk1992}
H.~L. Berk, J.~W. Van~Dam, Z.~Guo, and D.~M. Lindberg.
\newblock {Continuum damping of low‐n toroidicity‐induced shear Alfvén
  eigenmodes}.
\newblock {\em Physics of Fluids B: Plasma Physics}, 4(7):1806--1835, 07 1992.

\bibitem{Betti1992}
R.~Betti and J.~P. Freidberg.
\newblock {Stability of Alfvén gap modes in burning plasmas}.
\newblock {\em Physics of Fluids B: Plasma Physics}, 4(6):1465--1474, 06 1992.

\bibitem{Mett1992}
R.~R. Mett and S.~M. Mahajan.
\newblock {Kinetic theory of toroidicity‐induced Alfvén eigenmodes}.
\newblock {\em Physics of Fluids B: Plasma Physics}, 4(9):2885--2893, 09 1992.

\bibitem{Fulop1996}
T~Fülöp, M~Lisak, Ya~I Kolesnichenko, and D~Anderson.
\newblock Finite orbit width stabilizing effect on toroidal alfvén eigenmodes
  excited by passing and trapped energetic ions.
\newblock {\em Plasma Physics and Controlled Fusion}, 38(6):811, jun 1996.

\bibitem{Todo1995}
Y.~Todo, T.~Sato, K.~Watanabe, T.~H. Watanabe, and R.~Horiuchi.
\newblock {Magnetohydrodynamic Vlasov simulation of the toroidal Alfvén
  eigenmode}.
\newblock {\em Physics of Plasmas}, 2(7):2711--2716, 07 1995.

\bibitem{Todo1998}
Y.~Todo and T.~Sato.
\newblock {Linear and nonlinear particle-magnetohydrodynamic simulations of the
  toroidal Alfvén eigenmode}.
\newblock {\em Physics of Plasmas}, 5(5):1321--1327, 05 1998.

\bibitem{Fu2006}
G.~Y. Fu, W.~Park, H.~R. Strauss, J.~Breslau, J.~Chen, S.~Jardin, and L.~E.
  Sugiyama.
\newblock {Global hybrid simulations of energetic particle effects on the n=1
  mode in tokamaks: Internal kink and fishbone instability}.
\newblock {\em Physics of Plasmas}, 13(5):052517, 05 2006.

\bibitem{Wang2011}
X.~Wang, S.~Briguglio, L.~Chen, C.~Di~Troia, G.~Fogaccia, G.~Vlad, and
  F.~Zonca.
\newblock {An extended hybrid magnetohydrodynamics gyrokinetic model for
  numerical simulation of shear Alfvén waves in burning plasmas}.
\newblock {\em Physics of Plasmas}, 18(5):052504, 05 2011.

\bibitem{Chen2011}
Liu Chen and Fulvio Zonca.
\newblock Gyrokinetic theory of parametric decays of kinetic alfvén waves.
\newblock {\em Europhysics Letters}, 96(3):35001, oct 2011.

\bibitem{Hahm1995}
T.~S. Hahm and Liu Chen.
\newblock Nonlinear saturation of toroidal alfv\'en eigenmodes via ion compton
  scattering.
\newblock {\em Phys. Rev. Lett.}, 74:266--269, Jan 1995.

\bibitem{Chen2012}
Liu Chen and Fulvio Zonca.
\newblock Nonlinear excitations of zonal structures by toroidal alfv\'en
  eigenmodes.
\newblock {\em Phys. Rev. Lett.}, 109:145002, Oct 2012.

\bibitem{Qiu2023}
Z.~Qiu, L.~Chen, and F.~Zonca.
\newblock {Gyrokinetic theory of toroidal Alfvén eigenmode saturation via
  nonlinear wave–wave coupling}.
\newblock {\em Rev. Mod. Plasma Phys.}, 7(28), 2023.

\bibitem{GENE1}
F.~Jenko, W.~Dorland, M.~Kotschenreuther, and B.~N. Rogers.
\newblock Electron temperature gradient driven turbulence.
\newblock {\em Phys. Plasmas}, 7(5):1904--1910, 2000.

\bibitem{GENE2}
T.~G{\"o}rler, X.~Lapillonne, S.~Brunner, T.~Dannert, F.~Jenko, F.~Merz, and
  D.~Told.
\newblock The global version of the gyrokinetic turbulence code gene.
\newblock {\em J. Comput. Phys.}, 230(18):7053 -- 7071, 2011.

\bibitem{DiSiena2018}
A.~Di Siena, T.~Görler, H.~Doerk, E.~Poli, and R.~Bilato.
\newblock Fast-ion stabilization of tokamak plasma turbulence.
\newblock {\em Nuclear Fusion}, 58(5):054002, mar 2018.

\bibitem{DiSiena2019}
A.~Di~Siena, T.~Görler, E.~Poli, R.~Bilato, H.~Doerk, and A.~Zocco.
\newblock {Resonant interaction of energetic ions with bulk-ion plasma
  micro-turbulence}.
\newblock {\em Physics of Plasmas}, 26(5):052504, 05 2019.

\bibitem{Biancalani2021}
A~Biancalani, A~Bottino, A~Di Siena, Ö~Gürcan, T~Hayward-Schneider, F~Jenko,
  P~Lauber, A~Mishchenko, P~Morel, I~Novikau, F~Vannini, L~Villard, and
  A~Zocco.
\newblock Gyrokinetic investigation of alfvén instabilities in the presence of
  turbulence.
\newblock {\em Plasma Physics and Controlled Fusion}, 63(6):065009, 2021.

\bibitem{Beer1995}
M.~A. Beer, S.~C. Cowley, and G.~W. Hammett.
\newblock Field-aligned coordinates for nonlinear simulations of tokamak
  turbulence.
\newblock {\em Phys. Plasmas}, 2(7):2687--2700, 1995.

\bibitem{Scott1998}
B.~Scott.
\newblock Global consistency for thin flux tube treatments of toroidal
  geometry.
\newblock {\em Phys. Plasmas}, 5(6):2334--2339, 1998.

\bibitem{GoerlerPhD}
T.~{G{\"o}rler}.
\newblock {\em Multiscale effects in plasma microturbulence}.
\newblock PhD thesis, Universit{\"a}t Ulm, 2009.

\bibitem{Dimits2000}
A.~M. Dimits, G.~Bateman, M.~A. Beer, B.~I. Cohen, W.~Dorland, G.~W. Hammett,
  C.~Kim, J.~E. Kinsey, M.~Kotschenreuther, A.~H. Kritz, L.~L. Lao,
  J.~Mandrekas, W.~M. Nevins, S.~E. Parker, A.~J. Redd, D.~E. Shumaker,
  R.~Sydora, and J.~Weiland.
\newblock Comparisons and physics basis of tokamak transport models and
  turbulence simulations.
\newblock {\em Phys. Plasmas}, 7(3):969--983, 2000.

\bibitem{Lapillonne2009}
X.~Lapillonne, S.~Brunner, T.~Dannert, S.~Jolliet, A.~Marinoni, L.~Villard,
  T.~G{\"o}rler, F.~Jenko, and F.~Merz.
\newblock Clarifications to the limitations of the s-alpha equilibrium model
  for gyrokinetic computations of turbulence.
\newblock {\em Phys. Plasmas}, 16(3):032308, 2009.

\bibitem{Pueschel2010_2}
M.J. Pueschel, T.~Dannert, and F.~Jenko.
\newblock On the role of numerical dissipation in gyrokinetic vlasov
  simulations of plasma microturbulence.
\newblock {\em Computer Physics Communications}, 181(8):1428--1437, 2010.

\bibitem{Lapillonne2011_2}
X.~Lapillonne, B.~F. McMillan, T.~Görler, S.~Brunner, T.~Dannert, F.~Jenko,
  F.~Merz, and L.~Villard.
\newblock Nonlinear quasisteady state benchmark of global gyrokinetic codes.
\newblock {\em Physics of Plasmas}, 17(11):112321, 2010.

\bibitem{Heidbrink2008}
W.~W. Heidbrink.
\newblock {Basic physics of Alfvén instabilities driven by energetic particles
  in toroidally confined plasmasa)}.
\newblock {\em Physics of Plasmas}, 15(5):055501, 02 2008.

\bibitem{AjayCJPhD}
{Ajay, C. J.}
\newblock {\em Studying the effect of non-adiabatic passing electron dynamics
  on microturbulence self-interaction in fusion plasmas using gyrokinetic
  simulations}.
\newblock PhD thesis, {\'E}cole Polytechnique F{\'e}d{\'e}rale de Lausanne,
  2020.

\bibitem{Dominski2015}
J.~Dominski, S.~Brunner, T.~G{\"o}rler, F.~Jenko, D.~Told, and L.~Villard.
\newblock How non-adiabatic passing electron layers of linear
  microinstabilities affect turbulent transport.
\newblock {\em Phys. Plasmas}, 22(6):062303, 2015.

\bibitem{AjayCJ2020}
{Ajay C. J.}, S.~Brunner, B.~McMillan, J.~Ball, J.~Dominski, and G.~Merlo.
\newblock How eigenmode self-interaction affects zonal flows and convergence of
  tokamak core turbulence with toroidal system size.
\newblock {\em J. Plasma Phys.}, 86(5):905860504, 2020.

\bibitem{Navarro2011}
A.~Bañón~Navarro, P.~Morel, M.~Albrecht-Marc, D.~Carati, F.~Merz, T.~Görler,
  and F.~Jenko.
\newblock {Free energy balance in gyrokinetic turbulence}.
\newblock {\em Physics of Plasmas}, 18(9):092303, 09 2011.

\bibitem{AjayCJ2023}
Ajay C.J., B.F. McMillan, and M.J. Pueschel.
\newblock On the impact of temperature gradient flattening and system size on
  heat transport in microtearing turbulence.
\newblock {\em Nuclear Fusion}, 63(6):066024, apr 2023.

\bibitem{Lin1998}
Z.~Lin, T.~S. Hahm, W.~W. Lee, W.~M. Tang, and R.~B. White.
\newblock Turbulent transport reduction by zonal flows: Massively parallel
  simulations.
\newblock {\em Science}, 281(5384):1835--1837, 1998.

\bibitem{McMillan2019}
B.~F. McMillan, J.~Ball, and S.~Brunner.
\newblock Simulating background shear flow in local gyrokinetic simulations.
\newblock {\em Plasma Phys. Controlled Fusion}, 61(5):055006, mar 2019.

\bibitem{ToldPhD}
D.~Told.
\newblock {\em Gyrokinetic Microturbulence in Transport Barriers}.
\newblock PhD thesis, Universit{\"a}t Ulm, 2012.

\bibitem{Casson2009}
F.~J. Casson, A.~G. Peeters, Y.~Camenen, W.~A. Hornsby, A.~P. Snodin,
  D.~Strintzi, and G.~Szepesi.
\newblock Anomalous parallel momentum transport due to {E×B} flow shear in a
  tokamak plasma.
\newblock {\em Phys. Plasmas}, 16(9):092303, 2009.

\bibitem{WaltzTrans}
R.E. Waltz, R.L Dewar, and X.~Garbet.
\newblock Theory and simulation of rotational shear stabilization of
  turbulence.
\newblock {\em Physics of Plasmas}, 5:1784, 1998.

\bibitem{Dagnelie2019}
V.~I. Dagnelie, J.~Citrin, F.~Jenko, M.~J. Pueschel, T.~Görler, D.~Told, and
  H.~Doerk.
\newblock {Growth rates of ITG modes in the presence of flow shear}.
\newblock {\em Physics of Plasmas}, 26(1):012502, 01 2019.

\end{thebibliography}
\bibliographystyle{unsrt}

\end{document}